# Enhancing Model Fit Evaluation in SEM: Practical Tips for Optimizing Chi-Square Tests

Bang Quan Zheng†
*University of Arizona*

Peter M. Bentler
*UCLA*

Abstract

This paper aims to advocate for a balanced approach to model fit evaluation in structural equation modeling (SEM). The ongoing debate surrounding chi-square test statistics and fit indices has been characterized by ambiguity and controversy. Despite the acknowledged limitations of relying solely on the chi-square test, its careful application can enhance its effectiveness in evaluating model fit and specification. To illustrate this point, we present three common scenarios relevant to social and behavioral science research using Monte Carlo simulations, where fit indices may inadequately address concerns regarding goodness-of-fit, while the chi-square statistic can offer valuable insights. Our recommendation is to report both the chi-square test and fit indices, prioritizing precise model specification to ensure the reliability of model fit indicators.

[Word Count: 3,688]

†Correspondence should be addressed to Bang Quan Zheng, School of Government & Public Policy, University of Arizona, Tucson, AZ85719. E-mail: bangquan@arizona.edu.

## 1 Introduction

Social and behavioral scientists often grapple with complicated and abstract concepts such as democracy, value, ideology, identity, trust, and political tolerance, among others (Goren 2005; Sullivan et al. 1981; Davidov 2009; Acock et al. 1985; Feldman 1988). They often utilize Structural Equation Modeling (SEM) with latent variables, such as confirmatory factor analysis, to estimate statistical models that amalgamate indicators, aiming to gauge the underlying latent concepts. SEM's appeal lies in its dual capability to assess hypotheses regarding the influences of latent and observable variables on other variables, while also enabling simultaneous modeling of measurement error (Yuan and Liu 2021).

The crux of SEM lies in assessing model fit, which heavily relies on chi-square test statistics and fit indices, such as the normed fit index (NFI), comparative fit index (CFI), Tucker-Lewis fit index (TLI), root mean square error approximation (RMSEA), etc. Yet, there are currently no clear guidelines for interpreting model fit integrating chi-square test statistics and fit indices. Scholars have debated the applicability of fit indices due to the absence of a unanimous consensus on which fit indices to employ. In this discourse, there are those who contend that fit indices might possess limited practical utility (Barrett 2007), stressing the singular interpretation of the chi-square statistic. They express concerns that fit indices could potentially lead researchers to assert adequacy for a model that is incorrectly specified (Stone 2021). Some advocate against relying solely on preset cutoff values for fit indices, as these can be deceptive and misused. This perspective also highlights the issue of "cherry-picking," whereby researchers selectively choose a fit index that conforms to their preconceived viewpoint, thereby supporting a poorly fitting model (Stone 2021; Jackson et al. 2009; Kline 2015).

While SEM has made significant advancements in recent decades, this paper does not delve into those methodological advancements. Rather, its aim is to raise awareness and underscore the importance of adopting a well-balanced approach when evaluating model fit, while also providing guidance on its implementation. The argument presented is that the difficulties linked to chi-square tests don't solely arise from their constraints, but frequently result from a lack of proper understanding about their appropriate application. To exemplify, we will delineate three noteworthy scenarios frequently encountered by social and behavioral scientists when applying SEM, where fit indices might fall short in adequately addressing goodness-of-fit concerns, while the chi-square statistic can be effectively utilized. Scenario 1 involves misspecification, where the model used in the analysis poorly fits the data. Scenario 2 pertains to small sample sizes. A limitation of asymptotics is its lack of consideration for a statistic's behavior in small samples. Scenario 3 addresses non-normal data, where the maximum likelihood (ML) approach may not be effective. By employing alternative methods like the Lagrange Multiplier (LM) test, reweighted least squares, and Satorra-Bentler scaled robust estimators, we can achieve more accurate assessments of model fit. Furthermore, a proper interpretation and understanding of chi-square results, alongside other fit indices, are crucial for drawing accurate conclusions from data analysis. Our contention is that fit indices should not be solely relied upon as a cutoff point for model assessment; instead, researchers are strongly encouraged to report both chi-square and fit indices, with a greater emphasis on correctly specifying the model to ensure the trustworthiness of fit indices as meaningful indicators of model fit.

## 2 Review of Chi-Square Test and Fit Indices

In SEM, the model's parameters are held in the observed variable's covariance matrix $\boldsymbol{\Sigma}$, which can be written as $\boldsymbol{\Sigma} = \boldsymbol{\Lambda\Phi\Lambda}' + \boldsymbol{\Psi}$. Here, $\boldsymbol{\Lambda}$ is a matrix of factor loadings, $\boldsymbol{\Phi}$ is a matrix of factor covariances, and $\boldsymbol{\Psi}$ represents unique scores' covariances. In SEM, the expected structure of the population covariance matrix $\boldsymbol{\Sigma}$ is denoted as $\boldsymbol{\Sigma}(\boldsymbol{\theta})$, where $\boldsymbol{\theta}$ includes free parameters. Given that the sample covariance matrix $\boldsymbol{S}$ is an unbiased estimator of $\boldsymbol{\Sigma}$, an objective function $F[\boldsymbol{\Sigma}(\boldsymbol{\theta}), \boldsymbol{S}]$ gauges the difference between $\boldsymbol{\Sigma}(\boldsymbol{\theta})$ and $\boldsymbol{S}$. Our aim is to find $\widehat{\boldsymbol{\theta}}$, the estimated value of $\boldsymbol{\theta}$ that minimizes $F[\boldsymbol{\Sigma}(\boldsymbol{\theta}), \boldsymbol{S}]$. This involves iterative nonlinear programming, where we begin with an initial guess $\theta_i \in \{\theta_1, \dots, \theta_q\}$ and iteratively generate a sequence until it converges to $\widehat{\boldsymbol{\theta}}$, assuming smoothness in the partial derivatives of $F[\boldsymbol{\Sigma}(\boldsymbol{\theta}), \boldsymbol{S}]$ with respect to $\boldsymbol{\theta}$.

In covariance structure analysis with multivariate normally distributed variables, the most common method for evaluating goodness-of-fit is maximum likelihood (ML) (Jöreskog 1969). Equation 1 fits the model implied covariance matrix $\boldsymbol{\Sigma}(\boldsymbol{\theta})$ to the sample covariance matrix $\boldsymbol{S}$ using the Wishart likelihood function.

$$F_{ML}(\theta) = \log|\boldsymbol{\Sigma}\,(\boldsymbol{\theta})| - \log|\boldsymbol{S}| + tr(\boldsymbol{S}\boldsymbol{\Sigma}(\boldsymbol{\theta})^{-1}) - p \quad (1)$$

$$\widehat{\boldsymbol{\theta}}_{ML} = argmin\, F_{ML}(\boldsymbol{\theta}) \quad (2)$$

As shown in equation 2, at the minimum of the fit function $F_{ML}(\boldsymbol{\theta})$, $\widehat{\boldsymbol{\theta}}_{ML}$ contains parameter estimates $\widehat{\boldsymbol{\Lambda}}$, $\widehat{\boldsymbol{\Phi}}$, and $\widehat{\boldsymbol{\Psi}}$, where $\widehat{\boldsymbol{\Lambda}}$ is a matrix of estimated factor loadings, $\widehat{\boldsymbol{\Phi}}$ gives estimated factor covariances, and $\widehat{\boldsymbol{\Psi}}$ is the covariance matrix of error variables.

Moreover, the goodness-of-fit test statistic is defined by

$$T_{ML} = (N-1)F_{ML}(\widehat{\boldsymbol{\theta}}), \quad (3)$$

where $T_{ML}$ represents a test statistic calculated using $F_{ML}(\theta)$ based on the final parameter estimates and $N$ is the sample size. As $N \rightarrow \infty$, $T_{ML}$ is referred to a chi-square distribution with degrees of freedom $df = p^* - q$, where $p^* = p(p+1)/2$ and $q$ is the number of free parameters.

## *2.1 Limitations of Chi-Square Test*

Like any statistical test, the chi-square test has its limitations. For social science research, the major limitations are under-specification. Fitting a structural equation model often necessitates a larger number of items. However, unlike in psychology, social science research typically faces limitations in item availability, leading to model under-specification. Consequently, the chi-square test statistics tend to be substantially larger than the degrees of freedom ($df$), resulting in $p$-values approaching zero.

Second, the issue of statistical power, which refers to the ability to reject the null hypothesis, becomes particularly relevant when dealing with large sample sizes in SEM. Social science studies often involve relatively larger sample sizes. In SEM, the null hypothesis is that the covariance matrix $\Sigma$ is equal to the model-implied covariance matrix $\Sigma(\theta)$. To establish a plausible structural relationship, a $p$-value of 0.05 or greater is typically required, which is in contrast to regression analysis, where we expect a $p$-value of 0.05 or less to reject the null hypothesis. Yet, owing to the properties of the chi-square test as sample size increases, the likelihood of rejecting the null hypothesis strengthens. That is, in larger samples, the model's statistical power becomes large, potentially leading to null hypothesis rejection even when the model possesses minor inaccuracies. Fit indices have been developed to provide alternative measures of model fit.

Third, with small samples, $T_{ML}$ may not be reliable. Based on the assumption of multivariate normality, ML method provides the most widely used estimator in SEM (Hu et al. 1992; Jöreskog 1969; Bollen 1989). The behavior of this statistic is based on asymptotic properties, that is, $N$ must be sufficiently large. Previous research has found that a small sample $N$ is the main contributor to failure of asymptotic theory, but a large number of variables $p$ and/or parameters $q$, a small number of indicator loadings per factor, and small ratio of $N$ to $df$ also contribute to spurious goodness-of-fit model rejections (Arruda and Bentler 2017; Yuan and Bentler 1999; Zheng and Bentler 2023, 2021).

Fourth, $T_{ML}$ does not work for non-normal data. To follow a chi-square distribution ML is predicated on the assumption of multivariate normality, that is, normally distributed data. Nevertheless, in real-world data analysis, violations of these assumptions are common occurrences. Attempting to fit non-normal data using ML can result in the rejection of the null hypothesis, even when the model is correct. Owing to these limitations, the interpretability of $T_{ML}$ can be compromised. To address this issue, researchers have formulated a set of fit indices to enhance the evaluation of models.

## *2.2 Fit Indices*

Among the array of fit indices, the NFI, CFI, TLI, and RMSEA emerge as the most commonly used measures for assessing model fit. All fit indices but RMESA build upon a series of nested models, $M_i, \cdots, M_j, \cdots, M_k, \cdots, M_s$, spanning from the most constrained to the least constrained (the saturated model). Correspondingly, their associated chi-square test statistics are $T_i, \cdots, T_j, \cdots, T_k, \cdots, T_s$. Each index contributes its own set of metrics and limitations to the evaluation process, but all quantify the fit of the proposed model compared to the fit of the most constrained model. Values close to 1.0 are ideal. Bentler and Bonett (1980) introduced normed fit index (NFI), which is defined as:

$$NFI = \frac{T_i - T_k}{T_i}, \quad (4)$$

where $T_i$ represents the baseline chi-square value of an independence model, and $T_k$ represents the chi-square value of the particular model. Both $T_i$ and $T_k$ are derived from a specific fitting function, such as ML. When $T_i = T_k$, NFI=1, signifying a perfect fit. Conversely, when $T_k$ deviates from the expected values, NFI $< 1$. Bentler (1990) also proposed CFI, which is an incremental fit index that also compares the fit of the hypothesized model with that of a baseline model,

$$CFI = 1 - \frac{\lambda_k}{\lambda_i}, \quad (5)$$

where $\lambda_k$ and $\lambda_i$ are population noncentral parameters of hypothesized model and baseline models in practice estimated by difference between the test statistic and its degrees of freedom. Their sizes can be considered as indictors of model misspecification (Bentler 1990). Like CFI, TLI is defined as:

$$TLI = 1 - \frac{T_k/df_k}{T_i/df_i}, \quad (6)$$

where $T_k$ and $T_i$ are the same as in NFI, $df_k$ and $df_i$ are their respective degrees of freedom. Finally, an absolute fit index is given by

$$RMSEA = \sqrt{\max\left(\frac{T_k - df_k}{n \cdot df_k}, 0\right)}. \quad (7)$$

Whose size quantifies the extent of lack of fit. The better the fit, the closer the RMSEA value is to zero. The NFI has been found to be sensitive to the influence of small sample sizes, whereas variations in sample sizes have minimal impact on the CFI and TLI. Similarly, RMSEA, like NFI, is influenced by both model complexity and sample size. When the $df$ increases while $N$ decreases, RMSEA values tend to rise (Bentler 2006; Kenny et al. 2015). A simulation study by Hu and Bentler (1999) delved into the impact of various cutoff values for RMSEA, CFI, and TLI on rejection rates within accurate and misspecified models. Their findings suggested that, in general, a model can be deemed to exhibit relatively good fit when the RMSEA falls below 0.06, and both the CFI and TLI surpass 0.95.

A limitation of these fit indices lies in their strong reliance on metrics developed under the ML estimator (Xia and Yang 2019). Furthermore, if the chi-square test statistic encounters issues, the fit indices relying on them may also face challenges (Bentler 2006). In this context, the primary factor for obtaining dependable model fit evaluations for both chi-square tests and fit indices is ensuring the accurate specification of models.

## 3 Empirical Strategy and Simulation

This section delves into three key scenarios relevant to social science research where fit indices might lack in sufficiently tackling goodness-of-fit concerns. We present suggestions to optimize the utilization of the chi-square statistic, enabling its effective application in addressing these challenges. We conducted Monte Carlo simulations across varying sample sizes to visually illustrate their performances. To achieve this, we begin by establishing a population model from which we draw samples. Specifically, we opt for a confirmatory factor model represented as $\boldsymbol{X_i} = \boldsymbol{\Lambda\xi_i} + \boldsymbol{\varepsilon_i}$, where $\boldsymbol{X_i} = (X_{i1}, X_{i2}, \dots, X_{ip})'$ is a vector of $p$ observations on person $i$ in a population, and $i = 1, 2, \dots, N$. Under standard assumptions, this formulation leads to the covariance structure $\boldsymbol{\Sigma} = \boldsymbol{\Lambda\Phi\Lambda'} + \boldsymbol{\Psi}$. We chose a 3-factor model, with each factor being measured by 5 indicators, resulting in a total of 15 indicators. This model entails 33 free parameters and 87 $df$. The factor loading $\boldsymbol{\Lambda'}$ and $\boldsymbol{\Phi}$ are defined as:

$$\boldsymbol{\Lambda}' = \begin{bmatrix} .7 & .7 & .75 & .8 & .8 & 0 & 0 & 0 & 0 & 0 & 0 & 0 & 0 & 0 & 0 \\ 0 & 0 & 0 & 0 & 0 & .7 & .7 & .75 & .8 & .8 & 0 & 0 & 0 & 0 & 0 \\ 0 & 0 & 0 & 0 & 0 & 0 & 0 & 0 & 0 & 0 & .7 & .7 & .75 & .8 & .8 \end{bmatrix},$$

$$\boldsymbol{\Phi} = \begin{bmatrix} 1 & & \\ .3 & 1 & \\ .4 & .5 & 1 \end{bmatrix}$$

The data generating process consists of two steps. For a given *N*, a sample $\boldsymbol{\xi_i}$ is drawn from a covariance matrix $\boldsymbol{\Phi}$, while the unique factors $\boldsymbol{\epsilon_i}$ are drawn from a multivariate normal distribution with covariance $\boldsymbol{\Psi}$. For multivariate normal data, $\boldsymbol{\xi_i} = \boldsymbol{\Phi}^{1/2}\boldsymbol{Z}_{\xi}$ and $\boldsymbol{\varepsilon_i} = \boldsymbol{\Psi}^{1/2}\boldsymbol{Z}_{\varepsilon}$ where $\boldsymbol{\Phi}^{1/2}\boldsymbol{\Phi}^{1/2} = \boldsymbol{\Phi}$, $\boldsymbol{\Psi}^{1/2}\boldsymbol{\Psi}^{1/2} = \boldsymbol{\Psi}$, and both $\boldsymbol{Z}_{\xi}$ and $\boldsymbol{Z}_{\varepsilon}$ followed a standard normal distribution $\mathcal{N}(0, 1)$. This process was replicated 500 times across a range of sample sizes from 50 to 10,000. The performance of both the chi-square test and fit indices was computed over the 500 repetitions for each sample size.

### *3.1 Misspecified Model*

When the model is misspecified, $T_{ML}$ may not work well. In such cases, we can use the LM test for specification search and make appropriate modifications to the model accordingly. To illustrate, we need to assess the performance of $T_{ML}$ in cases where the models are incorrectly specified and compare it with a modified model based on the LM test.

To generate a misspecified model we modified the population model by adding an extra parameter to factors one and two respectively and set the factor loadings at the values of .3 and .2 respectively. The analysis model remains no change. Thus, the new factor loading matrix is defined as:

$$\widetilde{\boldsymbol{\Lambda}}' = \begin{bmatrix} 0.7 & 0.7 & 0.75 & 0.8 & 0.8 & 0 & 0 & 0 & 0 & 0 & 0 & 0.3 & 0 & 0 & 0 \\ 0.2 & 0.2 & 0 & 0 & 0 & 0.7 & 0.7 & 0.75 & 0.8 & 0.8 & 0 & 0 & 0 & 0 & 0 \\ 0 & 0 & 0 & 0 & 0 & 0.3 & 0 & 0 & 0 & 0 & 0.7 & 0.7 & 0.75 & 0.8 & 0.8 \end{bmatrix}.$$

Besides model misspecification, large sample sizes can also be a challenge for social science research. As the sample size increases, the tendency is for any model structure null hypothesis to face rejection. Consequently, many researchers find the large chi-square values to lack meaningful interpretation, prompting them to rely on fit indices for statistical justification in such scenarios. However, we contend that even when dealing with a large sample size, it is crucial to examine both chi-square and the LM test, along with fit indices.

Consider the following scenario: Initially, we have a model with a chi-square value of 900, based on a $N = 3000$ and $df = 5$. This result is deemed rejectable, indicating a poor fit for the model. However, the LM test suggests that adding an additional parameter might be beneficial. Acting on this advice, we introduce the extra parameter, resulting in a revised chi-square value of 600 with $df = 6$. Although the fit is still not satisfactory, the model's chi-square has dropped by 300 points for just 1 $df$, representing a highly significant improvement. This improvement through the addition of the extra parameter might hold valuable insights or meaningful relationships that were previously unidentified. Without this step, we would never have known about this potential improvement.

### *3.2 Small Samples*

Earlier research has noted that the ML estimator tends to exhibit an elevated rate of null hypothesis rejection in scenarios involving small sample sizes. Addressing this, the reweighted least squares (RLS) estimator has proven to be the most effective choice for such cases. The foundation of RLS is rooted in the normal-distribution GLS function initially introduced by Browne (1974),

$$F_{GLS} = 2^{-1}\, tr[\{(\boldsymbol{S} - \boldsymbol{\Sigma}(\boldsymbol{\theta}))\boldsymbol{V}\}^2]\,, \tag{8}$$

where $\boldsymbol{V}$ is a consistent estimator of $\boldsymbol{\Sigma}^{-1}$. In practice, $\boldsymbol{V} = \boldsymbol{S}^{-1}$ . To obtain the RLS function we first need compute the ML estimator $\widehat{\boldsymbol{\theta}}_{ML}$ (See equation 2) and the associated $\widehat{\boldsymbol{\Sigma}}_{ML}$. Then, also using (8),

$$T_{RLS} = \frac{n}{2}\, tr\{(\boldsymbol{S} - \widehat{\boldsymbol{\Sigma}}_{ML})\, \widehat{\boldsymbol{\Sigma}}_{ML}^{-1}\}^2. \quad (9)$$

Hence, the estimator is ML, but the GLS function (8) is evaluated with weight matrix $\boldsymbol{V} = \widehat{\boldsymbol{\Sigma}}_{ML}^{-1}$. Simulations show that RLS outperforms ML across varied sample sizes within the framework of a normal distribution (Zheng and Bentler 2023). An open-source R package for fitting the proposed models, 'RLS', can be found on GitHub: install_github("bzheng/RLS").

### *3.3 Non-Normal Data (Elliptical Distributions)*

Non-normal data are common in real-world social science data analysis, and $T_{ML}$ is not appropriate for situations where data deviate from normal distribution. As a solution, we can employ robust estimators. As an example, we have chosen the Satorra-Bentler scaled test. This estimator is specifically designed to address issues related to non-normality. For illustration purposes, we can generate simulations of non-normal data following an elliptical distribution, which represents symmetric distributions with heavy tails, based on the original population model. In the elliptical distribution condition $\boldsymbol{\xi_i} = r\boldsymbol{\Phi}^{1/2}\boldsymbol{Z}_{\xi}$, and $\boldsymbol{\varepsilon} = r\boldsymbol{\Psi}^{1/2}\boldsymbol{Z}_{\varepsilon}$ with r ~ $(3/\chi_5^2)^{1/2}$, $\boldsymbol{\Phi} = cov(\xi)$ and $\boldsymbol{\Psi} = cov(\boldsymbol{\varepsilon})$.

## 4 Monte Carlo Simulation Results

Figure 1. Chi-square Statistics and Fit Indices

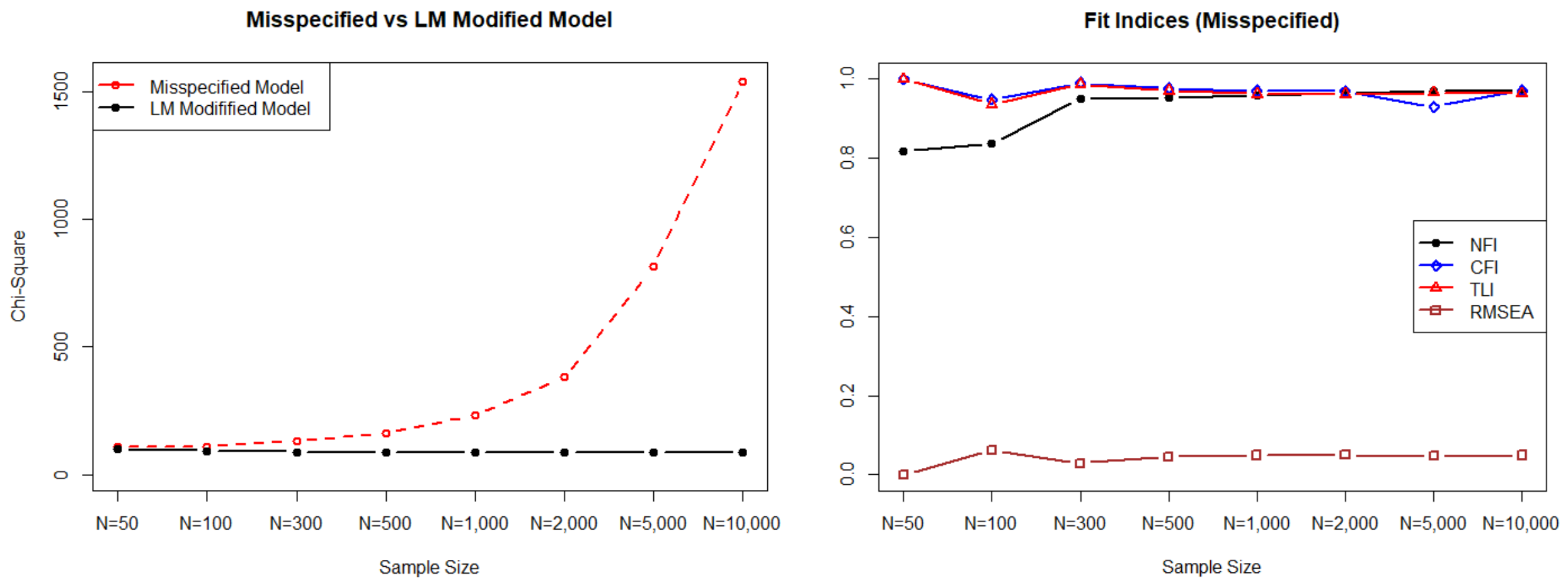

Figure 1 displays chi-square test statistics and fit indices for both a misspecified model and a LM modified model across various sample sizes. The chi-square test statistics of the misspecified model rise as the sample size N increases, indicating poor fit, while the LM modified model consistently demonstrates good fit. However, the fit indices in the right panel inaccurately suggest a good fit.

Figure 2. Chi-square Statistics of Non-normal data and Fit Indices

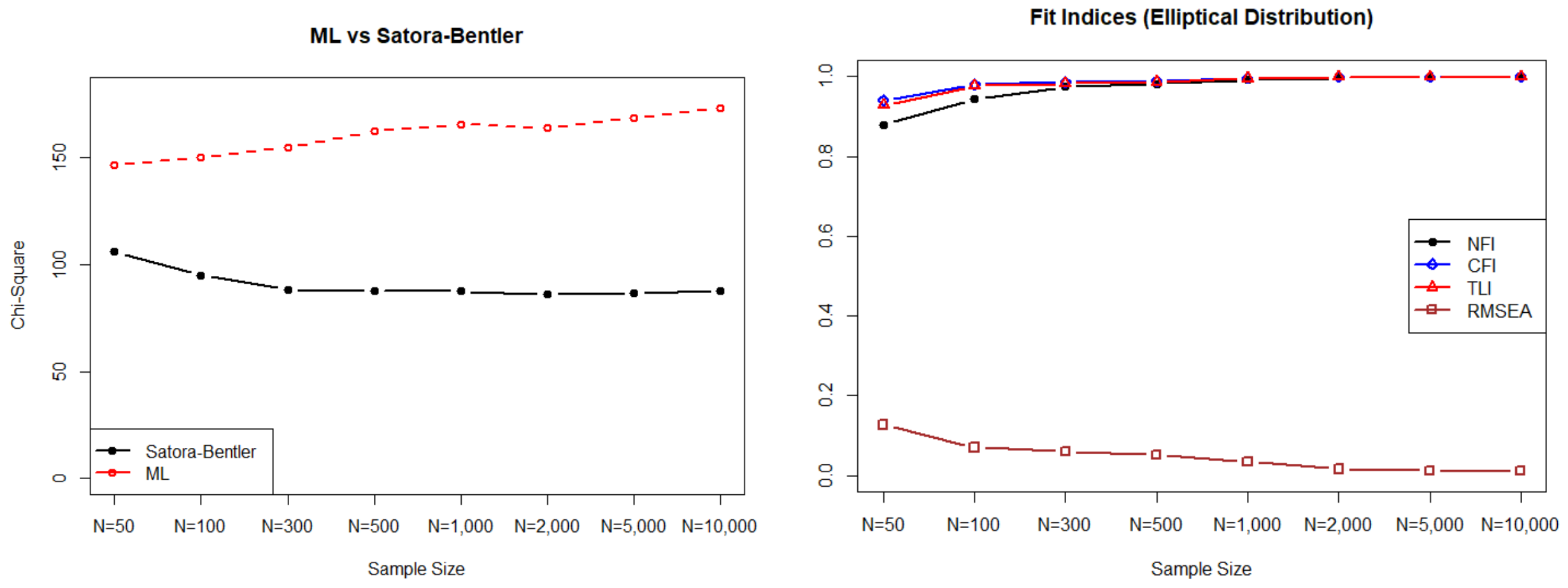

Figure 2 displays non-normal data (elliptical distribution). $T_{ML}$ deviated from the anticipated value of 87, and the fit indices in the right panel incorrectly demonstrated a good fit. Nonetheless, this concern can be rectified by utilizing a suitable robust estimator. In this instance, we employed the Satorra-Bentler scaled test. Notably, as Figure 3 shows, the chi-square test statistics rooted in Satorra-Bentler scaled chi-square maintain remarkable stability across nearly all sample sizes.

Figure 3. Chi-square Statistics on Small Samples and Fit Indices

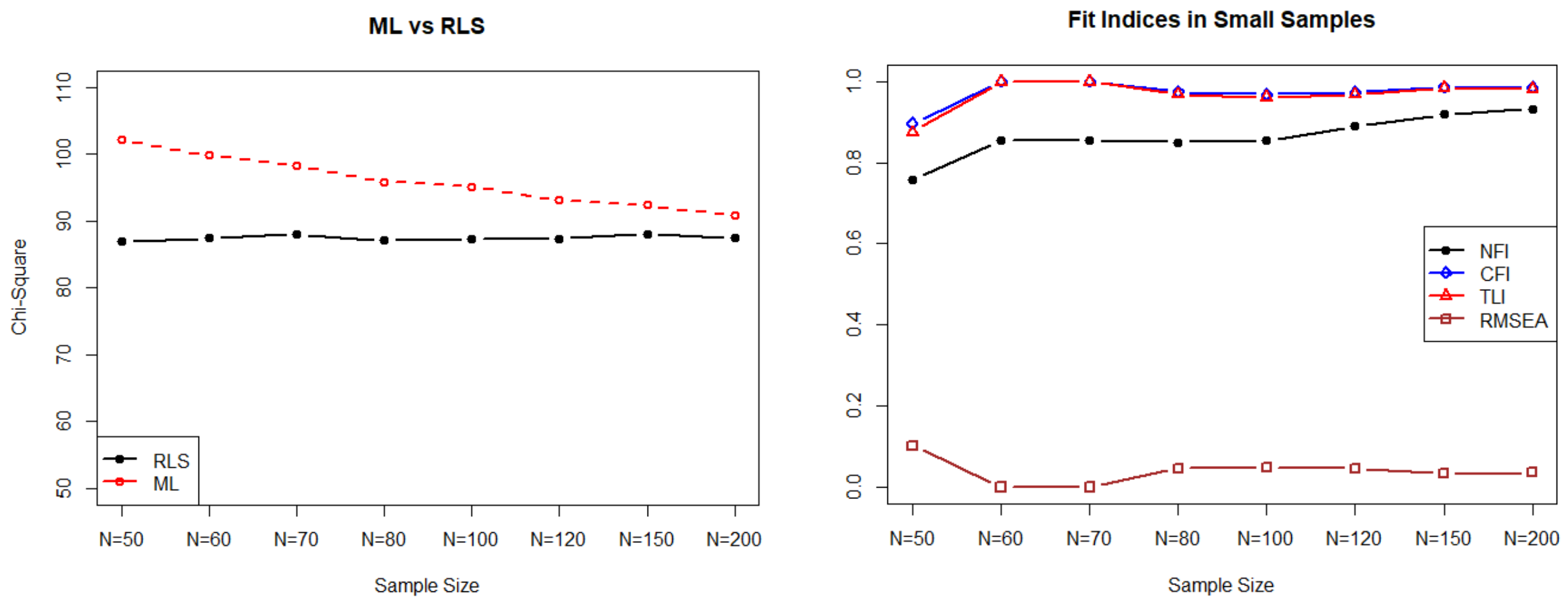

Figure 3 illustrates that when sample sizes are small (N<200), the default ML estimator becomes ineffective. In such instances, an alternative estimator is necessary for achieving precise outcomes. Figure 3 also highlights that with small sample sizes, $T_{ML}$ exceeds the expected value of 87, and the fit indices on the right panel also inaccurately showed a good fit. However, employing the RLS method yields notably consistent chi-square test statistics, resulting in enhanced accuracy and reliability in such scenarios.

## 5 Discussion and Conclusion

SEM entails numerous subtleties in statistical interpretations, accentuating the intricate nature of this methodology. In our study employing a series of Monte Carlo simulations, we have demonstrated that chi-square test statistics provide insights beyond those from the measurement of goodness-of-fit, encompassing vital insights into the overall performance and reliability of the model. Fit indices do not provide the same level of comprehensive evaluation as an appropriate chi-square test. Each fit index carries its distinct theoretical and analytical emphasis, limiting its ability to offer a holistic assessment of model fit. As such, relying solely on fit indices may obscure important misspecifications that could be of significant interest.

In sum, the chi-square test serves as a fundamental indicator of the model's overall fit quality and its compatibility with the observed data. Therefore, we advocate for the inclusion of chi-square test statistics in research reports, alongside fit indices, to present a more comprehensive and robust evaluation of the SEM model's fit. This approach allows for a balanced interpretation, taking into account both the nuanced insights provided by fit indices and the fundamental assessment offered by the chi-square test statistics.